\documentstyle[11pt,newpasp,twoside,psfig]{article}

\def\edcomment#1{\iffalse\marginpar{\raggedright\sl#1\/}\else\relax\fi}
\reversemarginpar

\begin{document}
\title{The case for late-time optical bumps in GRB afterglows
as a supernova signature}
 \author{J. S. Bloom}
\affil{Harvard-Smithsonian Center for Astrophysics, MC 20,
60 Garden Street, Cambridge, MA 02138, USA}

\begin{abstract}
Observations of late-time optical bumps have been reported for several
GRBs. The timescale for such bumps, and colors of such when available,
find a natural explanation as due to associated
supernovae. Ground-based and HST observations of the afterglow and
bump of GRB 011121, in particular, place the strongest constraints yet
on the physical nature of the supernova and any alternative
explanations, such as supranovae or dust echoes. I summarize the
search for underlying bumps in other GRBs and make the case for the
supernova hypothesis in light of observed bumps and bump
non-detections (e.g., GRB 010921). 
\end{abstract}

\noindent There is a good deal of theoretical motivation to expect to see bumps
in GRB afterglow lightcurves\footnotemark\footnotetext{For the
purposes of this presentation, I define a {\it bump} as: an increase
in flux above an extrapolated/interpolated light curve; extra source
of emission, owing, as reckoned, to processes of a different physical
mechanism. In Bloom et al.~(1999) we referred to the optical bump in
GRB\,980326 as a {\it rebrightening}. Bumps have been seen on shorter
timescales (e.g.~a few days 970508) but here, we focus on the
late-time bumps ($\sim$20--40 day timescale).}. For massive star
progenitors (Woosley 1993), a bump can arise either by
reflection/reprocessing of the afterglow light by surrounding dust
(Waxman \& Draine 2000; Esin \& Blandford 2000) or from a supernova that
accompanies the GRB. A bump might also arise from delayed energy
injection by the central source (Dai \& Lu 1998). If the afterglow
encounters a shell of material with higher density at $\sim10^{17}$
cm, then the afterglow could rebrighten. For compact
binary mergers (e.g., double neutron star coalescence), a late-time
bump is not a natural expectation since, a) mergers should occur in
homogeneous regions of low density, b) explosive nucleosynthesis
leading to a supernova is not expected around such systems, and c) the
creation of a stable neutron star after merger that is capable of
re-injecting energy after $\sim$20 days seems implausible. D.~Lazzati
(this workshop) has given an excellent overview of various progenitor
scenarios.

Thus the mere existence of bumps offer a strong discriminator between
merger and massive star scenarios. Within the confines of the massive
star progenitors model, the details of the bumps and accompanying
afterglows should also offer insight into the specifics of the
progenitors. For example, a bump from a supernova should have the
spectral and temporal characteristics of other supernovae observed in
the local universe; Woosley (1993) suggested that baryon contamination
of the relativistic jets could be minimized if this supernova was of
type Ic. A bump from thermal dust emission could have similar rise and
fall timescales as a supernova, but the spectrum would be thermal and
featureless, without any of the metal-line blanketing seen in
supernovae. A bump from dust echoes could mimic supernovae timescales
(Reichart 2001), but the bump should fade as a power-law rather than
an exponential.

\begin{figure}[p]
\centerline{
\psfig{figure=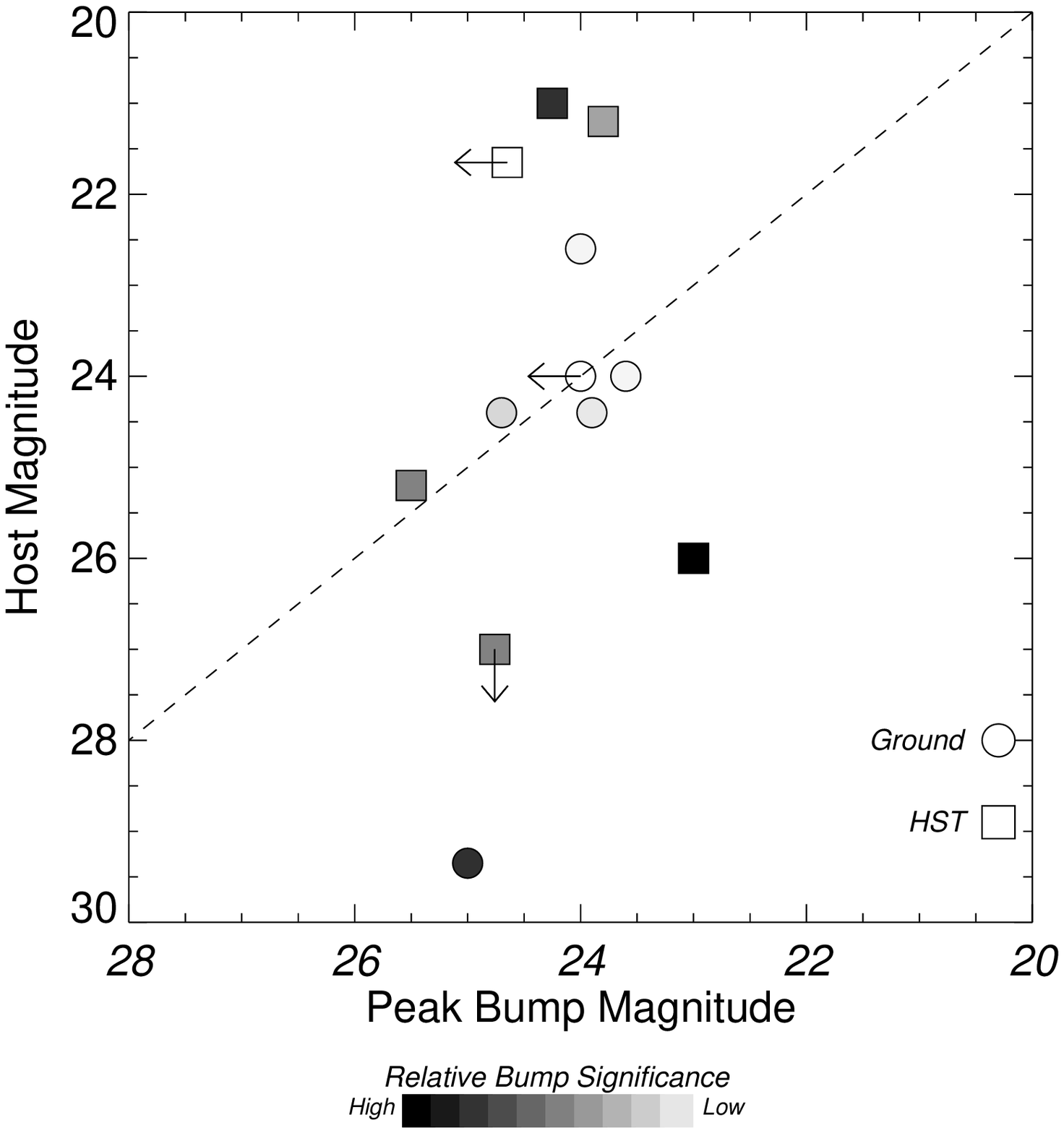,width=9cm,angle=0}}
\centerline{
\psfig{figure=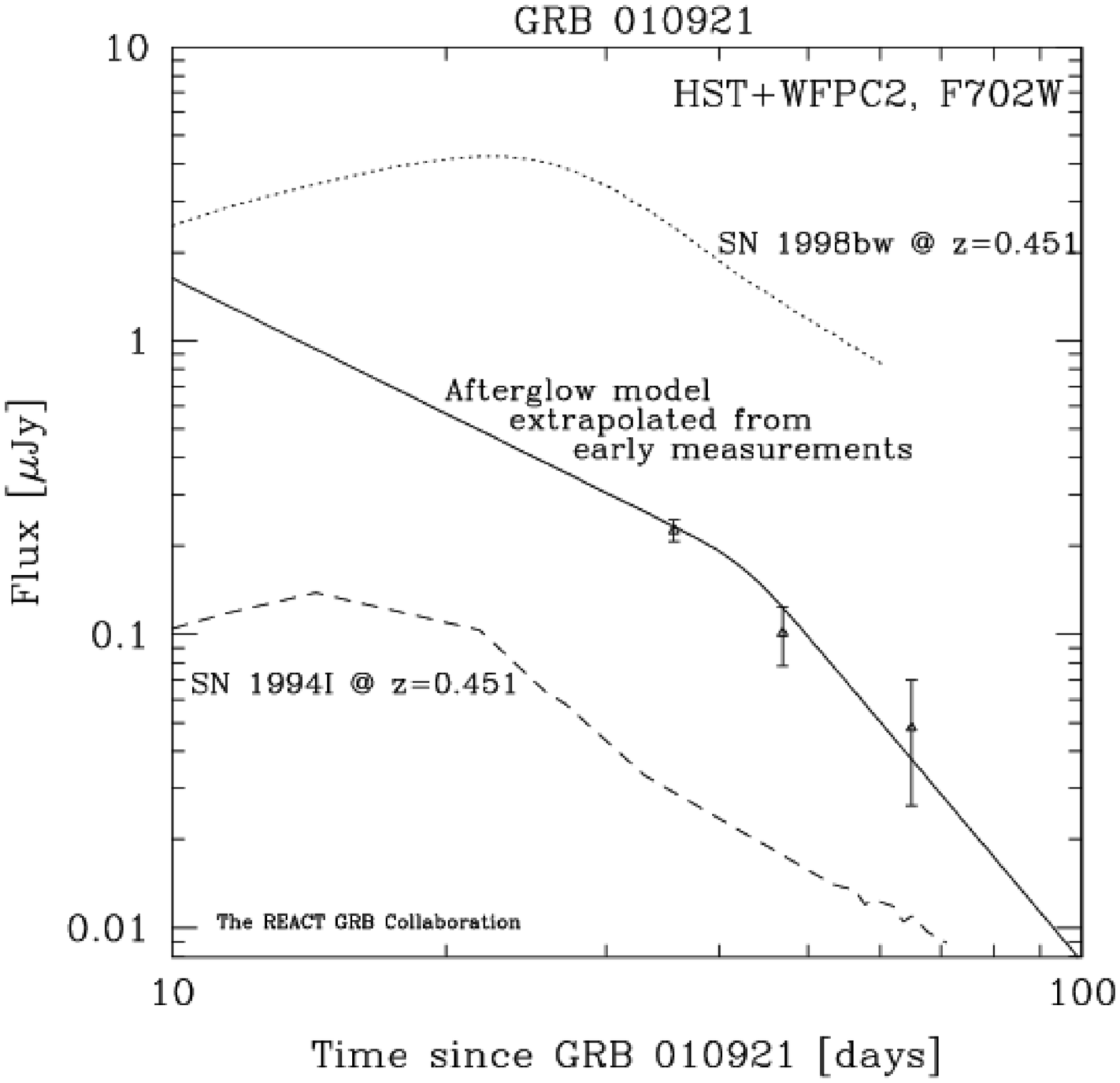,width=9cm,angle=0}}
\vskip -0.55cm
\caption[]{{\small (top) Peak bump magnitude versus host magnitude for those 
12 bursts with $z < 1.2$ where a bump search has been
conducted. Circles (squares) mark those bursts for which a bump was
searched for using ground-based (HST) data. Color gradations are
intended to show the relative significance of the detection. Note that
the significance of ground detection is stronger for bumps brighter
than the host and no bumps have been seen much fainter than the
host. Bumps found with HST can probe significantly fainter. (bottom)
The deepest non-detection of a bump with HST (adapted from Price et
al.~2002). Though any bump must be fainter than 1998bw at the redshift
of GRB 010921, it could still have been brighter than SN 1994I.}}

\end{figure}

\section{Bumps Associated With Cosmological GRBs}

The first observational detection of a bump associated with a
``cosmological'' burst\footnotemark\footnotetext{The detection of a
low-redshift supernova, SN 1998bw, coincident with a GRB (980425)
(Galama et al. 1998; E.~Pian, this workshop) provided an observational
link that until then had only been explored theoretically.  However,
since the burst must have been extraordinarily dim when compared to
other well-studied bursts---rather than definitively prove that
long-duration bursts arise in the core-collapse of a massive
stars---it is now believed that GRB 980425 might simply represent a
sub-class with a different bursting mechanism than the lion's share of
long-duration bursts (Bloom et al., 1998; Tan, Matzner, \& McKee,
2001). Nevertheless, we might consider SN 1998bw as the ultimate bump
associated with a GRB.} came with afterglow observations of GRB\, 980326
(Bloom et al.~1999). There, the early afterglow was seen to fade
rapidly in the first few days ($f_\nu \propto t^{-2}$) with a fairly
generic afterglow spectrum ($f_\nu \propto
\nu^{-0.8}$). At day 22 and day 28, imaging and spectroscopic
detections revealed a source that was about 60 times brighter than the
early extrapolation and significantly more red ($f_\nu \propto
\nu^{-2.3}$). By the next series of observations at 200 days, the bump
had faded by at least a factor of ten. HST imaging observations later
revealed a faint galaxy at the position of the transient (Fruchter et
al.~2001).

What powered the bump in GRB 980326? The timescale for rise and decay
are natural in the supernova interpretation. Moreover, the red color
of the bump can be understood as due to metal line blanketing of a
core-collapsed supernova redward of $\sim$4000 \AA\ in the restframe.
That the spectrum differed significantly from the early afterglow
spectrum suggests a different physical emission process than in the
afterglow. In other words, the bump in 980326 disfavors a synchrotron
shock origin, such as might be expected from a more dense external
medium or from delayed energy injection of a central source.

Bumps, attributed to SNe, have since been claimed in a number of other
cosmological bursts.  The bump in the afterglow of GRB 970228 was
based on observations in at least three bandpasses (Reichart 1999;
Galama et al.~2000). Moreover, the probable turn over in the spectrum
at $1\mu$m (Reichart 2001) could not be easily explained by thermal
emission from dust or a dust echo. The peak of redshifted thermal
emission should occur redward of $\sim$2 $\mu$m and dust echo spectra
should not exhibit a strong roll over. Optical/IR bumps discovered
20--30 days after GRBs 000911 (Lazzati et al.~2001), 990712 (Bj{\"
o}rnsson et al., 2001), 980703 (Holland et al.~2001) provided
suggestive evidence of bumps --- all of which the authors claimed
found a reasonable explanation in the context of an added supernova
component. For several bursts, however, no bump was detected to limits
of comparable brightness to 1998bw (see Price et al.~2002 for a
review). Note that due to severe line blanketing in the restframe UV,
a bump from a supernovae is not expected to be seen at optical
wavelengths for bursts beyond redshifts of $z \sim 1.2$ whereas bumps
from dust echoes could be seen to higher redshifts. Consistent with
the SN hypothesis, no bumps were detected by our group in the
afterglows of GRBs 010222 ($z=2.09$) and 000926 ($z=1.48$).

\section{Using HST to overcome observational impediments}

The ability to detect the bump in 980326 was aided by two important
occurrences. First, the rapid fading of the early afterglow
effectively removed any possible light contamination from the
afterglow beyond a few days. Second, the host galaxy was exceedingly
faint ($R=29.4$ mag).  Therefore, as we pointed out in Bloom et
al.~(1999), a bump in new GRBs could be easily outshone for bright
host galaxies and/or by bright afterglows. The latter difficulty can
be minimized by observing bursts with rapidly fading afterglow or with
early temporal breaks (presumably due to jetting). Even for these,
without an excellent multi-wavelength characterization of the
early-time afterglow, which allows for accurate predictions of the
flux at 10--40 days, bump detections will be more difficult to claim
unambiguously.  Overcoming the former host difficulty requires either
an intrinsically faint galaxy (like 980326) or the ability to resolve
the GRB afterglow from its host. To dig deeper into the bump
luminosity function, observations with the {\it Hubble Space
Telescope}, then, seemed the natural choice.
\begin{figure}[tb]
\centerline{\psfig{figure=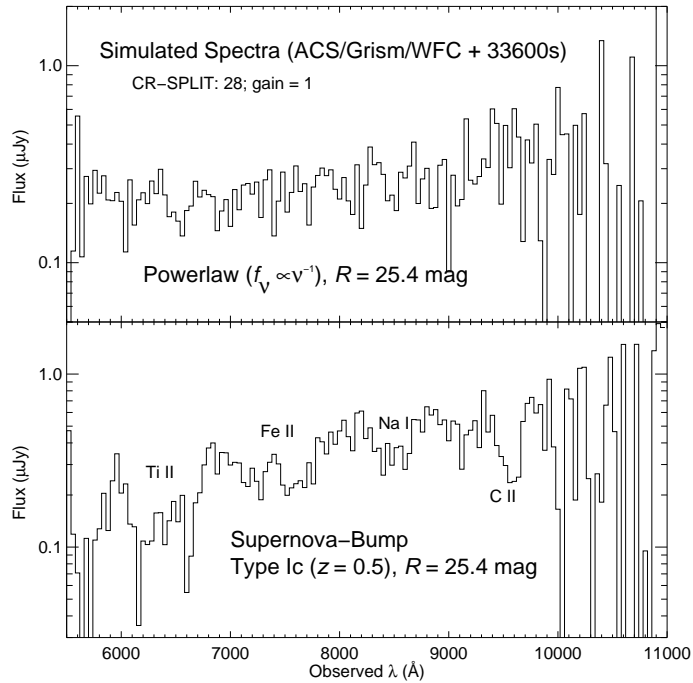,width=9.0cm,angle=0}}
\caption[]{A simulated SN spectrum at $R=25.4$ mag and $z=0.5$ as
observed with the {\it Advanced Camera for Surveys}+WFC on {\it HST}.
The template is SN 1994I (Type Ic), but dimmed to be 10 times fainter
than 1998bw at a redshift of $z=0.5$. Some broad core-collapsed SNe
spectral features are detectable even at such faint magnitude
levels. For nearer bursts or those with a brighter SN component (such
as we have already observed with GRB\,011121), the signal--to--noise
could be high enough to measure an expansion velocity of the
supernova.}
\end{figure}

As of this workshop there have been 12 reported searches for bumps in
GRB afterglows for bursts originating from $z < 1.2$.  Of these, half
utilized only ground-based observations, and half used both
ground-based and HST photometry. GRB 020405 (Price, this workshop) and
GRB 020331 (Soderberg, this workshop) are two new positive detections
from our HST program.  As illustrated in Figure 1, for the bursts
where a bump was seen or an upper-limit found from ground-based
observations alone, the host galaxy is a dominant
contaminant. Interestingly, only three bumps appear brighter than
their respective hosts at the same wavelengths.

HST observations have allowed us to probe nearly four magnitudes
fainter than the total magnitude of the host (in the case of GRB
010921; $z=0.45$). The non-detection of a bump in GRB 010921 is the
deepest constraint on a bump: any bump must have been at least 70\%
fainter than 1998bw at the redshift of the burst. In other words,
despite the added depth with HST photometry, {\it we still have not been
able to probe for bumps fainter than about} $M_V = -18$
{\it mag}.

\section{GRB 011121: The Best Case for a Bump}

Given the low redshift of GRB 011121 and extensive ground-based
observations at early times, we undertook a multi-epoch
multi-wavelength program with HST to try to detect a bump and test our
hypothesis that any detected bump could be due to a
supernova. Garnavich et al.~(2002) first pointed out that the
$R$-band flux at day 14 was higher than the extrapolated light curve
from earlier times. As described in more detail in Bloom et
al.~(2002), we detected a significant bump in four HST filters. In
that paper we described how the light curve and spectra over the next
78 days appeared to resemble a supernova. In addition, afterglow
models of 011121 provided the first clear-cut case of a
wind-stratified medium around the burst. This, coupled with the SN
interpretation, provides strong evidence for a massive star origin.

As both we and Garnavich et al.~(2003) described, the behavior of
the bump of 011121 did not follow a simple redshifted 1998bw: this
rise time appeared a to be quicker (17\% more rapid) and the peak flux
lower (55\% fainter). A few explanations could be possible. The
supernova could be of a different type than 1998bw (e.g., type IIn;
Garnavich et al., 2003). Instead, the supernova could be a type Ic but
with an energy and ejecta mass between that of 1998bw and 1994I (see
Figure 1). Alternatively, the supernova could be a dimmed version of
1998bw but with an explosion date that preceded the GRB by $-6 \pm 5$
days ($\chi^2$/dof $< 1$; Zeh et al., this workshop).

Accepting the supernova interpretation, the Zeh analysis is an
important one because it places constraints on the relative timing of
the supernova and the GRB. In a ``supranova'', a supramassive neutron
star is created as the massive star explodes to produce a supernova
(Vietri \& Stella 1998). When this neutron star spins down (via
magnetic breaking or gravitational radiation) --- losing centrifugal
support --- it collapses to form a black hole producing the GRB. Thus
a supernova could precede a GRB in time. Unfortunately, the time
offset is not a strong prediction of the supranova hypothesis---in
fact, we point out that the original incarnation of the model
suggested a 10 year offset, which is clearly ruled out by the 011121
observations.

\section{Conclusions}

The short time-offset in 011121 (consistent with contemporaneous GRB
and supernova explosions) underscores one of the more fundamental
observational limits in understanding the origin of bumps and GRB
progenitors. Since the explosion time of even the best studied
supernovae cannot be dated to better than about $\pm 2$ days, we may
never be able to infer the true time offset for short-delay
supranovae. That is, we may never be able to distinguish between the
collapsar model (zero delay) and short-delay supranova scenarios based
upon optical observations alone (neutrino arrival times would help!).

Clearly not all bumps can be fit with redshifted versions of 1998bw,
suggesting an intrinsic diversity in the properties of bumps (even
without appealing to a supernova interpretation). If these bumps are
due to supernovae then this diversity is unsurprising: local examples
of core-collapsed supernovae are anything but standard.  Mazzili et
al.~(2002) compiled spectra and light curves of a few well-observed
core-collapse supernovae and emphasized the large scatter in peak
fluxes ($M_V({\rm peak}) \approx -15$ to $-20$ mag) and broad range of
colors at peak.  Since the best non-detection of a bump reaches only
$\approx -18$ mag, I suggest that we may be only detecting the tip of
the iceberg of supernovae associated with GRBs.

There are several points to take away from the existing sample:
\begin{enumerate}
\item {\bf Late-time bumps are real and common.} Bumps have been 
detected with high significance in at least five GRB afterglows
(980326, 970228, 020405, 011121, 020331) with several other proposed
bump identifications (e.g., 000911, 990712, 980703). The
non-detections or ambiguity in the significance of bumps have been
largely due to contaminating light from the host galaxies. HST
observations have allowed us to push bump sensitivity almost four
magnitudes fainter than the integrated host magnitude. Still, the best
non-detection reaches only to $M_{V} \approx -18$ mag.
\item {\bf The best case suggests a supernova origin.} GRB 011121 is the best case for a bump and has a spectrum and light curve similar to a type Ic or IIn supernova. This
 is also the best example of a burst that occurred in a
 wind-stratified medium.
\item {\bf Limits on supranovae.} The original supranova hypothesis, which posited an offset in time between the GRB and the associated supernova of 10
years, is clearly ruled out. Modified-supranovae (with time offsets of
less than a $\sim$one week) are stilled allowed but may be
indistinguishable from collapsar models using bump observations alone.
\end{enumerate}
With such a small number of intensive bump searches, we know very
little about the frequency of occurrence of GRB bumps.  In a
supernova-centric interpretation, the current non-detections of bumps
at redshifts $z < 1.2$ may be due to a lack of sensitivity and
intrinsic supernovae diversity. Despite the concordance of a number of
detected bumps with simplistic {\it a priori} supernovae models, a
spectrum of a bump --- showing either the absence or presence of metal
lines --- would be most convincing one way or another. (Somewhat lost
in the various disagreements over the specifics of the bump emission
and timescales is: most viable alternatives put forth thus far to
explain bumps require a massive star progenitor rather than merger
products.) In the future, to more accurately characterize the nature
of bumps, we hope to undertake HST spectroscopy of a bump (see Figure
2) and determine if bump decay rates are indeed exponential.

\end{document}